# Structure and Magnetism of HoBaCo$_2$O$_5$ Layered Cobaltites with 0.02≤δ≤0.22

Lorenzo Malavasi,*[a] Yuri Diaz-Fernandez[a], M. Cristina Mozzati[b], Clemens Ritter[c]

[a] *Department of Physical Chemistry "M. Rolla", Viale Taramelli 16, Pavia, Italy.*
*Fax: +39 0382 987757; Tel: +39 0382 987757;*
*E-mail: lorenzo.malavasi@unipv.it*
[b] *Department of Physics "A. Volta", Via Bassi 6, Pavia, Italy.*
[c] *Institute Laue-Langevin, Boite Postale 156, F-38042, Grenoble, France.*

In this paper we have studied, by means of high-resolution neutron powder diffraction and magnetic susceptibility, the structural and magnetic features of selected samples of the HoBaCo$_2$O$_{5+\delta}$ layered cobaltite in the low oxygen content range (0.02≤δ≤0.22. The results shows a strong antiferromagnetic contribution at room temperature coupled to an intermediate spin state of the Co$^{3+}$ ions.

# Introduction

Recently there has been a growing interest in the study of layered cobaltites of general formula $RE$BaCo$_2$O$_{5+\delta}$ ($RE$=rare earth) due to their rich structural, electronic, and magnetic phase diagrams resulting from the strong coupling between charge, orbital, and spin degrees of freedom [1-5].

One of the most interesting aspect of layered cobaltites is the large oxygen non-stoichiometry they show, whit $\delta$ theoretically varying from 0 to 1. This degree of freedom allows a continuous doping of the square-lattice CoO$_2$ planes which not only influences the mean Co valence state but also the carrier nature, the Co spin-state and the bandwidth.

Among the most interesting properties of layered cobaltites, which are tuned by the oxygen content, Taskin *et al.* have shown a remarkable change in the charge carriers nature in GdBaCo$_2$O$_{5.\delta}$ and NdBaCo$_2$O$_{5.\delta}$ single crystals moving from $\delta \approx 0.40$ to $0.60$ allowing them to give a solid experimental support to the idea that strong electron correlations and spin-orbital degeneracy can bring about a large thermoelectric power in transition-metal oxides [6,7].

A study of the physical properties of layered cobaltites as a function of the oxygen content may revel interesting and new features on their properties. However this kind of study is experimentally quite demanding particularly in systems where fast oxygen diffusion occurs, as in the present case. In particular, the evolution of the magnetic structure along with the variation in the oxygen content looks to be a very interesting issue, particularly when related to the spin state of the Co$^{3+}$ ions, which can be found in the low-, intermediate- and high-spin state. For the Ho layered cobaltite the only available works as a function of $\delta$ are at $\delta$=0 and 0.5. For the first composition still some doubts are present related to the spin-state of the Co ions since magnetic studies on the HoBaCo$_2$O$_5$ sample appeared up to now could not discriminate between the two probable models involving $Co^{3+}$ ions either in HS or in IS state. The observed G-type AFM structure does not provide additional information since both models may be qualitatively explained by the GK rules for superexchange [8,9]. For the $HoBaCo_2O_{5.5}$ sample the magnetic structure is quite complicated and carefully investigated only recently [10]. From this work it has been observed an antiferromagnetic structure with a $2a_p \times 2a_p \times 4a_p$ magnetic unit cell containing four crystallographically independent Co ions, two octahedrally coordinated and two pyramidally coordinated. Of the two Co ions in the octahedra, one has been found to be in the high-spin (HS) state while the other in a mixed intermediate- (IS) and low-spin (LS) state. The pyramidally coordinate $Co^{3+}$ ions were found to be in the intermediate spin-state [10]. It was also observed that the complex magnetic structure of $HoBaCo_2O_{5.5}$ contains both positive and negative exchange interactions between nearest neighbours.

In this paper we are going to present room-temperature (RT) neutron diffraction (ND) data coupled to susceptibility measurements for samples with $0.02 \leq \delta \leq 0.22$ in order to show the nature of the magnetic coupling in this composition interval. These results are of great importance since may further shed light on the spin-state and magnetic coupling variation induced by the oxygen content change.

# Experimental

Powder samples of $HoBaCo_2O_{5+\delta}$ have been prepared by conventional solid state reaction from the proper stoichiometric amounts of $Ho_2O_3$, $Co_3O_4$, and $BaCO_3$ (all Aldrich ≥99.99%) by repeated grinding and firing for 24 h at 1050-1080 °C. $Ho_2O_3$ was first heated at 900°C overnight before being used in the reaction. Oxygen content was fixed according to thermogravimetry (TGA) measurements by annealing $HoBaCo_2O_{5+\delta}$ pellets at selected $T$ and $p(O_2)$ in a home-made apparatus for at least 48 hours followed by rapid quenching in liquid nitrogen [see details in Ref 11]. X-ray diffraction (XRD) patterns at room temperature were acquired on a "Bruker D8 Advanced" diffractometer equipped with a Cu anode in a $\theta$-$\theta$ geometry. Measurements were carried out in the angular range from 10 to 110° with 0.02° step-size and acquisition time for each step of at least 10 s. Diffraction patterns were refined by means of Rietveld method with the FULLPROFILE software [12]. Static magnetization was measured at 100 Oe from 360 K down to 2 K with a SQUID magnetometer (Quantum Design). Neutron data have been collected at room temperature on the D2B instrument at the ILL facility (Grenoble).

## Results and Discussion

Figure 1 and 2 show, respectively, the room temperature neutron and x-ray diffraction patterns for the three $HoBaCo_2O_{5+\delta}$ samples investigated here, i.e. with $\delta=0.02$, 0.08 and 0.22. X-ray diffraction is needed in order to properly determine the crystal structure of the compounds since small deviation of the crystal symmetry can not be revealed by neutron diffraction only. Sample with $\delta=0.02$ is characterized by an orthorhombic unit cell (space group, *Pmmm*) with $a \sim b \sim a_p$ (where $a_p$ represents the pseudo-cubic lattice parameter of the perovskite unit cell) and $c \sim 2a_p$. Table 1 reports the structural parameters for the various samples investigated in the present work. At $\delta=0.02$ the number of $Co^{3+}$ ions is slightly higher that $Co^{2+}$ ions and nearly all of these cobalt ions are coordinated in corner shared square base pyramids formed by the oxygen neighbours while the La and Ba atoms are ordered and form alternated layers along the *c*-axis. At this composition the oxygen ions in addition to $\delta=0$ populate the $HoO_\delta$ layer by distributing on the 1*b* Wycoff position, that is the (0,0,½) position. Only a small fraction of the $Co^{3+}$ ions (2%) are octahedrally coordinated. The orthorhombic distortion of the unit cell is extremely small as can be appreciated from the values of the lattice parameters reported in Table 1. However, a clear indication of this distortion was observed in some diffraction lines such as the (110) and (212) located at about 32° and 46°, respectively (see inset of Figure 2). The presence of an orthorhombic symmetry for a $HoBaCo_2O_{5+\delta}$ sample with an oxygen content above 5 was not yet reported in the current literature.

Samples with $\delta=0.08$ and 0.22 present a tetragonal symmetry (space group, *P4/mmm*) with $a=b=a_p$ and $c \sim 2a_p$. The unit cell parameters and cell volume are reported, as well, in Table 1. As can be appreciated, along with the increase of the oxygen content the cell volume decreases. The contraction of the cell volume is due to the oxidation of the $Co^{2+}$ ions to $Co^{3+}$ ions since, for the same coordination, the first one has a ionic radius of 0.885 Å (high-spin configuration) while the second has a ionic radius of 0.75 Å (high-spin configuration).

The refinement of the neutron diffraction pattern was carried out according to the structural starting-model obtained from the x-ray diffraction results. Inset of Figure 1 shows the Rietveld refined pattern for the sample with δ=0.02. As can be appreciated there are two distinct zones of the pattern where the refinement is not able to account for the intensity of several peaks. The extra-peaks not indexed with the refinement of the nuclear contribution only clearly comes from a magnetic contribution which, of course, can not be detected with the x-ray diffraction technique. This results is quite interesting because it shows that all the three samples investigated shows high-intensity magnetic peaks at room temperature.

The extra peaks can be indexed with a propagation vector (½ ½ 0) in the $a_p \times a_p \times 2a_p$ unit cell. This is the so-called G-type magnetic structure which is characterized by the Co ions being antiferromagnetically coupled with their six nearest Co neighbours along the three crystallographic axes. In order to consider this magnetic contribution we included into the refinement the magnetic cell. Figure 3 shows, as a selected examples, the refined pattern, considering both nuclear and magnetic contribution, for the δ=0.02 sample (i.e. the same of the inset of Figure 1). From the refinement of the magnetic structure we could observe that the magnetic moments lie in the *ab*-plane of the structure without any evidence of an out-of-plane contribution which may give origin to a ferromagnetic (FM) moment. The refined magnetic moment for the Co ions for this sample (in Bohr magnetons units) is 1.95(5). The values for all the three samples are reported in Table 1. As can be appreciated, along with the increase in the oxygen content the net magnetic moment per Co ions decreases. However, this piece of information needs to be coupled to the knowledge of the $Co^{3+}$ spin state.

The magnetic moment derived from neutron diffraction is not the same as the effective free ion moment value which is $\mu = 2\sqrt{S(S+1)}$ for a spin state *S*. The spin-only neutron scattering value for a localized model corresponds to the sum of the unpaired electrons and for a spin state $S$ it is $\mu = 2S$ [8].

We are aware that the magnetic moments calculated from the neutron data at room temperature are most probably not representative of their saturation values. For example, for δ=0 the magnetic moment was found to be ~ 2 $\mu_B$ at room temperature and saturates at about 200 K to ~2.5 $\mu_B$, i.e. at room temperature the value is about 80% of the saturation value. However, it is clear that the neutron data seems to rule out the presence of $Co^{3+}$ in the high-spin state. For the δ=0.02 sample the full spin only magnetic moment for 1:1 HS-$Co^{2+}$/ HS-$Co^{3+}$ should be greater than 3.5 $\mu_B$.

In order to give a stronger picture of the results found we carried out susceptibility measurements on the samples investigated here. Figure 4 reports the inverse magnetic susceptibility curves for the lowest and highest oxygen content investigated in this work, i.e. δ=0.02 and 0.22. From the fit of the curves at high temperature (i.e. up to room temperature) with the Curie-Weiss formula we could determined a $p_{eff}$ which is in perfect agreement with a IS for the $Co^{3+}$ ions. This result, coupled to the RT neutron diffraction data, allow us to conclude that the long-range AFM order found at room temperature for the $HoBaCo_2O_{5+\delta}$ samples with 0.02≤δ≤0.22 is realized between HS-$Co^{2+}$ and IS-$Co^{3+}$ ions.

## Conclusion

In this paper we have investigated the structural and magnetic properties of HoBaCo$_2$O$_{5+\delta}$ samples with 0.02≤δ≤0.22 by means of room temperature x-ray and neutron diffraction and susceptibility measurements. The data collected shows the presence of a long-range G-type AFM order in all the samples. The magnetic moment progressively decreases with the increase of δ as a consequence of the Co oxidation and not of any spin-state change. The coupled information gained from the refinement of the magnetic peaks in the neutron diffraction patterns and the susceptibility measurements allowed us to conclude that in this δ-range the AFM order is realized between HS-Co$^{2+}$ and IS-Co$^{3+}$ ions. Further neutron diffraction measurements as a function of temperature will be realized in order to possibly observe spin-state transitions at lower temperatures. The results presented here are the first ones collected on HoBaCo$_2$O$_{5+\delta}$ samples for δ≠0 or 0.5 and are of great significance in order to understand the correlation between the Co ions spin state and the oxygen content and may be also useful in order to re-consider the available data on the δ=0 composition.

# Acknowledgements

This work has been supported by the "Celle a combustibile ad elettroliti polimerici e ceramici: dimostrazione di sistemi e sviluppo di nuovi materiali" FISR Project of Italian MIUR. We recognize the support of the UNIPV-Regione Lombardia Project on Material Science and Biomedicine. ILL neutron facility and European Community financial support is acknowledged.

**Figure Caption**

**Fig. 1 -** Neutron patterns for $HoBaCo_2O_{5+\delta}$ at the three oxygen contents investigated in the present paper in a reduced 2θ-range in order to better highlight this part of the spectrum. Inset: Rietveld refined pattern for δ=0.02 sample with the nuclear part only.

**Fig. 2 -** X-ray diffraction patterns for $HoBaCo_2O_{5+\delta}$ at the three oxygen contents investigated in the present paper in a reduced 2θ-range in order to better highlight this part of the spectrum. Inset: selected 2θ-range of the main Figure.

**Figure 3** – Refined neutron pattern for $HoBaCo_2O_{5.02}$ (same sample of the inset of Figure 1) considering the nuclear contribution and the magnetic contribution coming from an AFM ordering.

**Figure 4 -** Inverse molar susceptibility of the $HoBaCo_2O_{5+\delta}$ for δ=0.02 and 0.22.

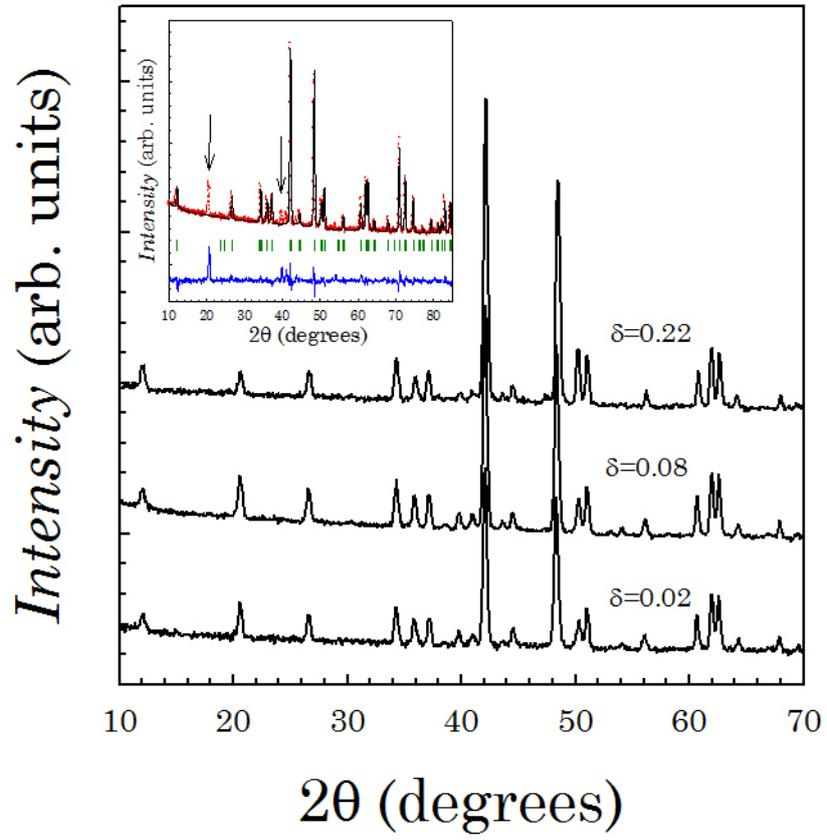

**Figure 2**

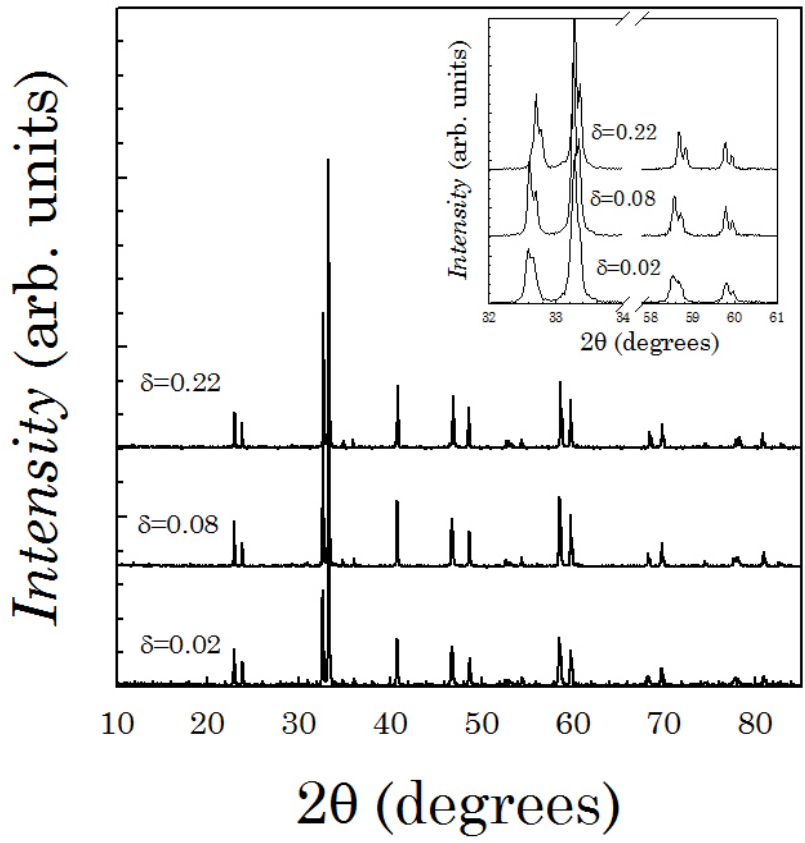

**Figure 3**

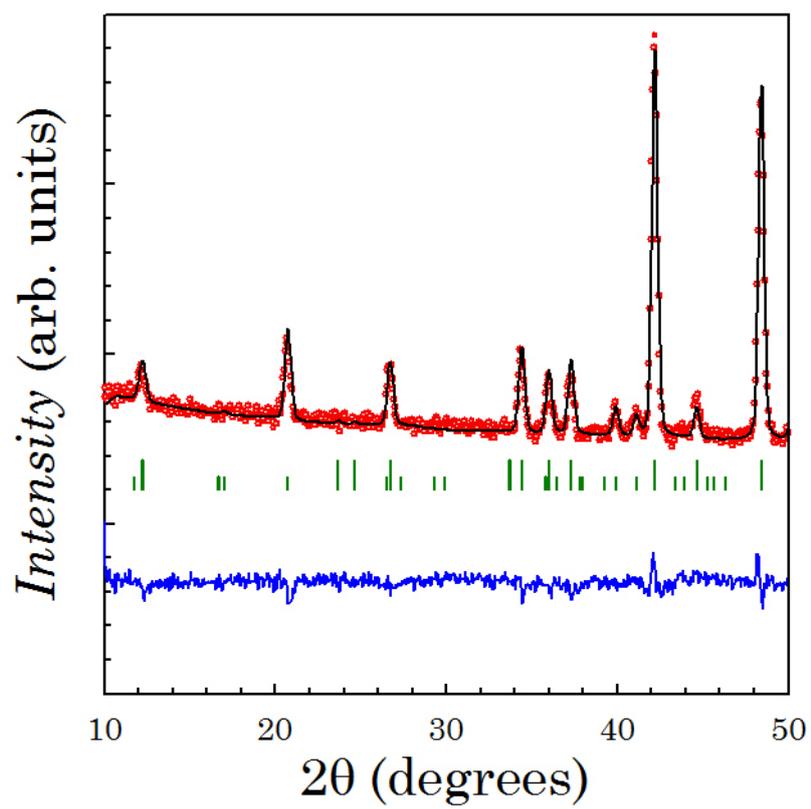

**Figure 4**

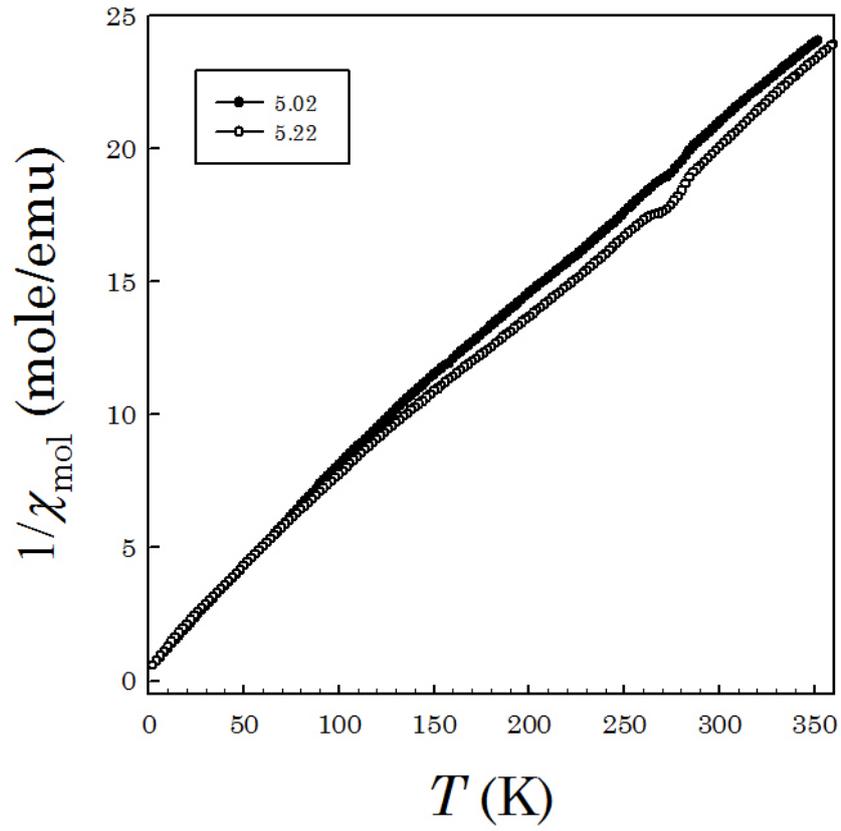

# Table

**Table 1** Structural parameters derived from the x-ray diffraction patterns at room temperature for the HoBaCo$_2$O$_{5+\delta}$ samples and refined magnetic moments from neutron diffraction.

|  | $\delta=0.02$ | $\delta=0.08$ | $\delta=0.22$ |
|---|---|---|---|
| $a$ (Å) | 3.8920(1) | 3.88621(7) | 3.87704(8) |
| $b$ (Å) | 3.8853(1) | 3.88621(7) | 3.87704(8) |
| $c$ (Å) | 7.4845(2) | 7.4870(1) | 7.4993(1) |
| $V$ (Å$^3$) | 113.179(5) | 113.074(4) | 112.725(4) |
| $\chi^2$ | 1.26 | 1.43 | 1.38 |
| $\mu_B$ | 1.97(5) | 1.95(5) | 1.36(5) |